\documentclass[%
 reprint,
showpacs,preprintnumbers,
 amsmath,amssymb,
 aps,
pra,
]{revtex4-1}
\usepackage{amsmath,amssymb,graphicx,color,bm,epstopdf}
\usepackage{dcolumn}
\usepackage{bm}

\begin{document}

\title{Band gaps induced by vacuum photons in closed semiconductor cavities}

\author{O.V. Kibis}\email{Oleg.Kibis@nstu.ru}

\affiliation{Department of Applied and Theoretical Physics,
Novosibirsk State Technical University, Novosibirsk 630073,
Russia}

\affiliation{Science Department, University of Iceland, Dunhaga 3,
IS-107, Reykjavik, Iceland}

\author{K.B. Arnardottir}

\affiliation{Science Department, University of Iceland, Dunhaga 3,
IS-107, Reykjavik, Iceland}

\affiliation{Division of Physics and Applied Physics, Nanyang
Technological University 637371, Singapore}

\author{I.A. Shelykh}

\affiliation{Science Department, University of Iceland, Dunhaga 3,
IS-107, Reykjavik, Iceland}

\affiliation{Division of Physics and Applied Physics, Nanyang
Technological University 637371, Singapore}



\begin{abstract}
We consider theoretically a closed (zero-dimensional)
semiconductor microcavity where confined vacuum photonic mode is
coupled to electrons in valence band of the semiconductor. It is
shown that vacuum-induced virtual electron transitions between
valence and conduction bands result in renormalization of electron
energy spectrum. As a consequence, vacuum-induced band gaps appear
within the valence band. Calculated values of the band gaps are of
sub-meV scale, that makes this QED effect to be measurable in
state-of-the-art experiments.
\end{abstract}

\maketitle

{\it Introduction.---} There are physical situations where the
electron-photon interaction cannot be considered as a weak
perturbation (so-called regime of strong light-matter coupling).
In this regime, it is necessary to consider the system ``electron
+ field" as a whole. Such a bound electron-photon object, which
was called ``electron dressed by field" (dressed electron), became
commonly used model in modern physics
\cite{Cohen-Tannoudji_b98,Scully_b01}. The interest in this field
is stimulated by the possibility of the achievement of the hybrid
--- half-light half-matter --- excitations which can demonstrate
peculiar properties. Therefore, the regime of strong light-matter
coupling was extensively investigated both theoretically and
experimentally in a variety of the systems, including optical
planar microcavities with semiconductor
\cite{Kasprzak,Balili,Yamamoto} and organic
\cite{KenaCohen,Plumhof,Mazzeo} quantum wells, microcavities with
individual quantum dots
\cite{QDStrongCouplingPillar,QDStrongCouplingDefect,QDStrongCouplingWGM}
and others. Effects of strong coupling can be used for variety of
technological applications {\cite{PolDevices}}, including novel
types of the lasers \cite{Christopoulos,Schneider}, optical
switches and logic gates \cite{Paraiso,Amo,Anton}, all-optical
integrated circuits \cite{Espinosa}, sources of entangled photon
pairs \cite{Johne} and others. Among most bright phenomena of
strong light-matter coupling, we have to note the field-induced
modification of energy spectrum of dressed electrons
--- also known as a dynamic (ac) Stark effect --- which was discovered
in atoms many years ago \cite{Autler_55} and has been studied in
details in various atomic and molecular systems
\cite{Cohen-Tannoudji_b98,Scully_b01}. {For} solids, the dynamic
Stark effect results in the gap opening within electron energy
bands \cite{Elesin_69,Galitskii_70,Schmitt_88,Vu_04,Kibis_11}.

In order to turn usual ``bare'' electrons into ``dressed''
electrons, a characteristic energy of electron-photon interaction
should be increased. This can be done in two different ways. The
first of them consists in using a strong laser-generated
electromagnetic field (the case of large photon occupation
numbers) \cite{Cohen-Tannoudji_b98,Scully_b01}. The second way
consists in decreasing {the} effective volume where
electron-photon interaction takes place. This can be realized with
embedding {an} electron system inside a microcavity
\cite{KavokinBook,PolDevices}. It should be stressed that there is
no formal physical difference between ``real'' photons (quanta of
electromagnetic wave) and ``virtual'' ones (quanta of vacuum
fluctuations inside a cavity). Therefore, the strong coupling of
electrons to vacuum photonic mode inside a cavity can open energy
gaps in various solids in the same way as an usual dynamic Stark
effect. To study this phenomenon, the new area of
interdesciplinary theoretical research at the border between
quantum electrodynamics and physics of semiconductors was opened
during last years. In the previous papers on the subject
\cite{Kibis_11_1,Kibis_13,Espinosa_14}, the vacuum-induced
modification of electron energy spectrum in solids was studies
exclusively for {\it opened} cavities, including both
two-dimensional cavities and one-dimensional ones. Unfortunately,
vacuum-induced gaps in these cavities are very small. As a
consequence, experimental observation of the gaps is very
difficult since the scattering of conduction electrons in real
solids washes the gaps. Therefore, it is necessary to find
physical objects where the discussed effects can be observable in
state-of-the-art experiments. In the given Brief Report, we will
demonstrate that the vacuum-induced gaps can be giant in {\it
closed} (zero-dimensional) cavities of macroscopically large size.

\begin{figure}
\centering
\includegraphics[width=0.4\textwidth]{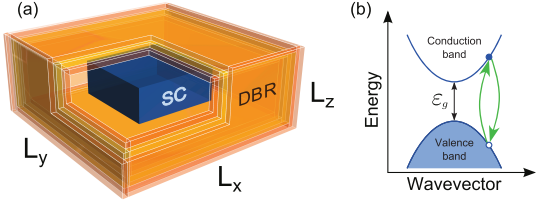}
\caption{(Color online) (a) Sketch of the system under
consideration: a bulk semiconductor (SC) embedded into a closed
(zero-dimensional) microcavity formed by distributed Bragg
reflectors (DBR) with dimensions $(L_x, L_y, L_z)$; (b) Electron
energy spectrum of the bulk semiconductor with vacuum-induced
(virtual) interband electron transitions marked by the
arrows.}\label{Figsketch}
\end{figure}

{\it Model. ---} Let us consider a semiconductor embedded inside a
closed (zero-dimensional) cavity [see Fig. \ref{Figsketch}(a)]. In
what follows, we will assume that the cavity dimensions
$(L_x,L_y,L_z)$ are macroscopically large as compared with
characteristic de-Brougle wave length of the electrons. Therefore,
the electron energy spectrum of conduction and valence bands of
the considered semiconductor sample are the same as for bulk
semiconductor,
$\varepsilon_{v,c}(\mathbf{k})=\pm{\varepsilon_g}/{2}\pm{\hbar^2k^2}/{2m_{v,c}}$,
where $\varepsilon_g$ is the semiconductor band gap, $m_{v,c}$ are
the electron effective masses in valence and conduction bands,
$\mathbf{k}$ is the electron wave vector. Interacting with vacuum
photonic mode of the cavity, valence electrons perform virtual
transitions between valence and conduction bands [see Fig.1(b)],
which modify the electron energy spectrum. In order to find the
renormalized spectrum of valence electrons, $\varepsilon$, let us
apply the conventional diagrammatic approach based on Green's
functions.

\begin{figure}
\centering
\includegraphics[width=0.35\textwidth]{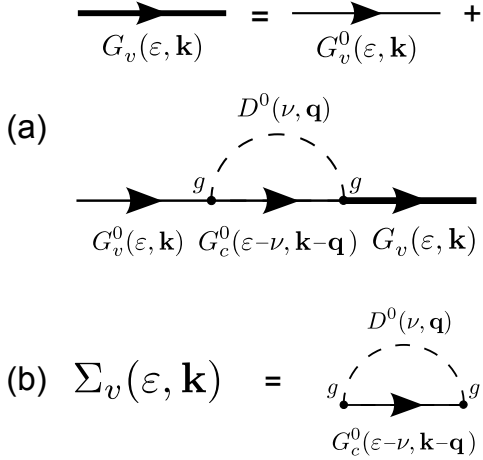}
\caption{(a) The Dyson equation for renormalized electron Green's
functions corresponding to the  valence band; (b) The self-energy
operator responsible for the photon-induced dressing of the
valence band. The dashed line corresponds to a virtual cavity
photon and $g$ denotes the electron-photon coupling
constant.}\label{feynman}
\end{figure}

Generally, a dressed electron in the valence band with the energy
$\varepsilon$ and the wave vector $\mathbf{k}$ is described by the
Dyson equation [see Fig.~2(a)] which can be written in the
algebraic form as
\begin{equation}
\label{Greenf}
G_v(\varepsilon,\textbf{k})=\frac{G_v^0(\varepsilon,\textbf{k})}{1-\Sigma_v(\varepsilon,\textbf{k})
G_v^0(\varepsilon,\textbf{k})},
\end{equation}
where the self-energy for valence electrons interacting with
cavity photons [see Fig.~2(b)] is defined by the expression
\begin{equation}
\label{Sigma}
i\Sigma_v(\varepsilon,\textbf{k})=\sum_{\textbf{k}^\prime}g^2(\textbf{k}^\prime,\textbf{k})
\int\frac{d\nu
}{2\pi}iD^0(\nu)iG_c^0(\varepsilon-\nu,\textbf{k}^\prime),
\end{equation}
$
G_c^0(\varepsilon,\textbf{k})=(\varepsilon-\varepsilon_c(\textbf{k})-i0)^{-1}
$ is the Green's function for bare conduction electron, $
G_v^0(\varepsilon,\textbf{k})=(\varepsilon-\varepsilon_v(\textbf{k})+i0)^{-1}
$
is the Green's function for bare valence electron,
\begin{equation}\label{D0}
D^0(\nu)= \frac{2\omega}{(\nu-\omega+i0)(\nu+\omega-i0)}
\end{equation}
is the photon Green's function, $\nu$ is the photon energy,
$\omega$ is the eigenenergy of cavity photons, and
$g(\textbf{k}^\prime,\textbf{k})$ is the electron-photon coupling
constant. The sought dispersion relation of valence electrons
dressed by vacuum fluctuations, $\varepsilon(\mathbf{k})$, is
given by the poles of the renormalized Green function
(\ref{Greenf}). These poles are defined by the equation
\begin{equation}\label{p}
1-\Sigma_v(\varepsilon,\textbf{k})
G_v^0(\varepsilon,\textbf{k})=0.
\end{equation}
As to the electron-photon coupling constant, within the dipole
approximation it can be written as
\begin{equation}\label{g}
g(\textbf{k}^\prime,\textbf{k})=\left|\left\langle\psi_c(\textbf{k}^\prime),0\left|\hat{\mathbf{d}}\hat{\mathbf{E}}\right|\psi_v(\textbf{k}),1\right\rangle\right|,
\end{equation}
where $\psi_{v,c}(\textbf{k})$ are electron wave functions in the
valence and conduction bands, the symbols $(0,1)$ describes the
photon occupation numbers of the cavity mode, $\hat{\mathbf{d}}$
is the operator of electric dipole moment,
\begin{equation}\label{E}
\hat{\mathbf{E}}=i\sqrt{\frac{\omega}{2\epsilon\epsilon_{0}}}
[\hat{a}^{\dagger}\mathbf{u}^{*}- \hat{a}\mathbf{u}]
\end{equation}
is the operator of electric field corresponding to the cavity
photon eigenmode $\mathbf{u}$,  and $\hat{a},\,\hat{a}^\dagger$
are the operators of annihilation and creation of photons in the
cavity eigenmode. In order to simplify the calculations, let us
restrict our consideration to the case of $L_z\ll L_{x,y}$. Then
eigenmodes of the cavity photons are described by the expression
\begin{equation}\label{u}
\mathbf{u}(\mathbf{r})\approx\mathbf{e}\sqrt{\frac{8}{V}}\sin\left({q_z
z}\right)\sin\left({q_x x}\right)\sin\left({q_y y}\right),
\end{equation}
where $\mathbf{r}=(x,y,z)$ is the radius-vector written in the
Cartesian coordinates, $V=L_xL_yL_z$ is the cavity volume,
$\mathbf{e}=(e_x,e_y)$ is the polarization vector of the cavity
eigenmode, $q_{x,y,z}=\pi l_{x,y,z}/L_{x,y,z}$ are characteristic
photon wave vectors of the eigenmode, and
$l_{x,y,z}=\pm1,\pm2,\pm3...$. Then, taking into account the
electron interaction with a ground photon mode of the cavity, we
can write eigenenergy of cavity photons as $\omega\approx{\hbar
c\pi}/{L_z}$. Substituting Eqs.~(\ref{E})--(\ref{u}) into the
Eq.~(\ref{g}), we get
\begin{equation}\label{gf}
g^2(\textbf{k}^\prime,\textbf{k})=\frac{|d_{cv}|^2}{16}\frac{\omega}{\epsilon\epsilon_0V}
\sum_{\mathbf{q}}\delta_{\mathbf{k}^\prime,\mathbf{k}-\mathbf{q}}
\end{equation}
where the photon wave vector of ground eigenmode is
\begin{equation}\label{q0}
\mathbf{q}=\left(\frac{\pi n_x}{L_x},\;\frac{\pi n_y}{L_y},\;
\frac{\pi n_z}{L_z}\right),
\end{equation}
$n_{x,y,z}=\pm1$, and $d_{cv}$ is the interband matrix element of
the dipole moment at $\mathbf{k}=0$. Within the {Kane} model of
semiconductor band structure \cite{Voon}, this matrix element can
be written as $|d_{cv}|\approx
|e|\hbar/\sqrt{{2m_c\;\varepsilon_g}}$.

{\it Results and discussion. ---} In order to find the energy
spectrum of dressed valence electrons, $\varepsilon$, we have to
solve Eq.~(\ref{p}). This equation can be solved analytically in
the approximation of negligibly small photon wave vector
$(q\approx0)$, when the coupling constant (\ref{gf}) takes the
simple form
\begin{equation}\label{gf22}
g^2(\textbf{k}^\prime,\textbf{k})\approx {\frac{\omega
|d_{cv}|^2}{2\epsilon\epsilon_0V}}\,\delta_{\textbf{k}^\prime,\textbf{k}}.
\end{equation}
Substituting the constant (\ref{gf22}) into Eq.~(\ref{Sigma}), we
arrive at the self-energy
\begin{equation}
\label{Sigma2}\begin{split}
&\Sigma_v(\varepsilon,\textbf{k})=i{\frac{\omega
|d_{cv}|^2}{2\epsilon\epsilon_0V}}\\
&\times\int\frac{d\nu
}{2\pi}\frac{2\omega}{(\nu-\omega+i0)(\nu+\omega-i0)(\varepsilon-\nu-\varepsilon_c(\mathbf{k})-i0)}.
\end{split}
\end{equation}
Performing contour integration in Eq.~(\ref{Sigma2}) over the
upper half-plane, $\rm{Im}(\nu)>0$, and applying the residue
theorem, we can write  the self-energy (\ref{Sigma2}) as
\begin{equation}
\label{Sigma3} \Sigma_v(\varepsilon,\textbf{k})={\frac{\omega
|d_{cv}|^2}{2\epsilon\epsilon_0V[\varepsilon+\omega-\varepsilon_c(\mathbf{k})]}}.
\end{equation}
Then Eq.~(\ref{p}) is
\begin{equation}\label{p2}
[\varepsilon+\omega-\varepsilon_c(\mathbf{k})][\varepsilon-\varepsilon_v(\mathbf{k})]
-\frac{\omega d^2_{cv}}{2\epsilon\epsilon_0V}=0
\end{equation}
and immediately gives the sought energy spectrum of the valence
band modified by vacuum fluctuations,
\begin{eqnarray}\label{En}
\varepsilon&=&\frac{\varepsilon_c(\mathbf{k})+\varepsilon_v(\mathbf{k})}{2}-\frac{\omega}{2}\nonumber\\
&-&\frac{\eta(\mathbf{k})}{2}\sqrt{[\omega-\varepsilon_c(\mathbf{k})+\varepsilon_v(\mathbf{k})]^2+4(\hbar\Omega_R)^2},
\end{eqnarray}
where
$\Omega_R={|d_{cv}|}\sqrt{{{\omega}}/{{2\epsilon\epsilon_0V\hbar^2}}}$
is the effective Rabi frequency of the considered electron-photon
system, and
\begin{equation}\label{eta}\nonumber
\eta(\mathbf{k})=\left\{\begin{array}{rl} -1,
&\omega>[\varepsilon_c(\mathbf{k})-\varepsilon_v(\mathbf{k})]\\\\
1, &\omega<[\varepsilon_c(\mathbf{k})-\varepsilon_v(\mathbf{k})]
\end{array}\right..
\end{equation}
It follows from Eq.~(\ref{En}) that vacuum fluctuations induce the
energy gap
\begin{equation}\label{Gap}
\Delta\varepsilon=2\hbar\Omega_R=|d_{cv}|\sqrt{\frac{2\omega}{\epsilon\epsilon_0V}}
\end{equation}
within the valence band at electron wave vectors, $\mathbf{k}$,
satisfying the resonant condition
$\omega=[\varepsilon_c(\mathbf{k})-\varepsilon_v(\mathbf{k})]$.
\begin{figure}
\begin{center}
\includegraphics[width=0.5\textwidth]{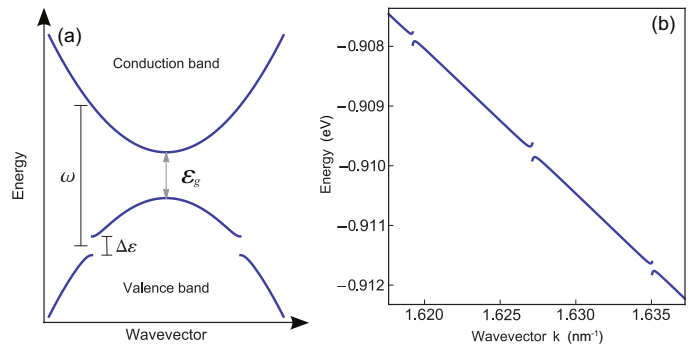}
\caption{(Color online) (a) Energy spectrum of dressed electrons
within the approximation of a negligible small photon wavevector;
(b) Exact energy spectrum of dressed valence electrons, calculated
numerically for GaAs microcavity with dimensions $L_z=100$ nm and
$L_x=L_y=500$ nm for the electron wave vector
$\mathbf{k}=(k_x,k_y,k_z)$ with the components
$k_x=k_y=k/\sqrt{2}$ and $k_z=0$.}\label{FigGap}
\end{center}
\end{figure}

The gapped energy spectrum (\ref{En}) is pictured schematically in
Fig.~3(a). Outside the approximation of $q\approx0$, we have to
solve Eq.~(\ref{p}) numerically with using exact expression for
the coupling constant (\ref{gf}). Due to finite value of the
photonic wavevector $\mathbf{q}$, the vacuum-induced gap acquires
fine structure: Instead of a single gap, several gaps appear [see
Fig.~3(b)]. Positions of the gaps in the $\mathbf{k}$ space are
defined by the resonant condition,
$\varepsilon_c(\mathbf{k}-\mathbf{q})-\varepsilon_v(\mathbf{k})=\omega$,
where the photon wave vector of {the} cavity eigenmode,
$\mathbf{q}$, is given by Eq.~(\ref{q0}).
\begin{table}
\center \caption{Vacuum-induced band gaps, $\Delta\varepsilon$,
calculated for various semiconductor cavities with dimensions
$L_x=L_y=5L_z$.}\label{Table}
\begin{tabular}{l|cccc}
\hline Semiconductor   &CdTe   &ZnO    &GaN    &GaAs\\ \hline
$L_z$ (nm)      &100    &85     &88     &100\\
$\Delta\varepsilon$ (meV) &0.36    &0.28   &0.27   &0.38\\\hline
\end{tabular}
\end{table}

In Tab.~\ref{Table}, we present values of vacuum-induced band gaps
which are calculated for various semiconductors using
Eq.~\eqref{Gap}. It is seen that the gaps are around meV and,
therefore, are several orders of magnitude larger compared to the
vacuum corrections of energies of individual atoms in the absence
of the cavity (for instance, the Lamb shift). It should be noted
also that the calculated values of the gap for closed
(zero-dimensional) cavities, given in Tab.~\ref{Table},
essentially exceed values of the vacuum-induced gaps appearing in
open (two-dimensional and one-dimensional) cavities
\cite{Kibis_11_1,Kibis_13,Espinosa_14}. Physically, the increasing
of gap values in closed (zero-dimensional) cavities as compared
with open ones arises from the strong (delta-function-like)
singularity in the density of photon states. In order to observe
the photon-induced gaps experimentally, angle-resolved
photoemission spectroscopy (ARPS) looks to be most appropriate.
Indeed, ultra-violet laser-based ARPS \cite{Zhang_13,Kiss_05}
provides sub-meV resolution which is enough for detecting gap
values given in Tab.~\ref{Table}. As to dissipation processes
arisen from nonideal structure of the cavity, they lead to the
finite spectral linewidth of the cavity mode, $\Gamma$ (see, e.g.,
Refs. \cite{Bajoni_08,Nelsen,Yoshie}). Correspondingly, the
photon-induced gaps, $\Delta\varepsilon$, are observable under the
condition of $\Delta\varepsilon>\Gamma$. This condition can be
easily satisfied for the gaps given in Tab.~\ref{Table} since
modern technologies allow to fabricate microcavities with the
spectral linewidth $\Gamma\sim10^{-5}$~eV \cite{Nelsen}. It should
be noted also that closed cavities of very high quality allow to
detect the vacuum Rabi splitting in individual quantum dots, which
is less than the gaps given in Tab.~\ref{Table} \cite{Yoshie}. As
a consequence, the predicted effect can be observed in
state-of-the-art experiments.

{\it Conclusion. ---} We calculated energy spectrum of valence
electrons in closed (zero-dimensional) semiconductor cavities,
which is modified by vacuum fluctuations of electromagnetic field.
The feature of the spectrum consists in vacuum-induced energy gaps
of sub-meV scale within valence band, which can be observed
experimentally. As a consequence, the discussed effect can open
new area of interdisciplinary experimental research where quantum
electrodynamics and physics of solids meet.

{\it Acknowledgements. ---} We thank O.~Kyriienko and
T.~Espinosa-Ortega for valuable discussions. This work is
supported by FP7 ITN ``NOTEDEV", Tier 1 project ``Polaritons for
novel device applications", FP7 IRSES projects POLATER, POLAPHEN
and QOCaN, RFBR projects 13-02-90600 and 14-02-00033. O.V.K.
thanks the University of Iceland for the hospitality and
acknowledges the support from the RANNIS project 141241-051.

\end{document}